\documentclass[twocolumn,prl]{revtex4-1}
\usepackage{natbib}
\usepackage{ulem}
\usepackage{graphicx, epsfig}
\usepackage{latexsym}
\usepackage{epsfig}
\usepackage{latexsym}
\usepackage{amstext}
\usepackage{ulem}
\usepackage{amssymb}
\usepackage{array}
\usepackage{color}
\usepackage{graphicx}
\usepackage{dcolumn}
\usepackage{amsmath}
\usepackage{amssymb}
\usepackage{amsfonts}
\usepackage{amssymb}
\usepackage{amsthm}
\usepackage{newlfont}
\usepackage{graphics}
\usepackage{mathrsfs}
\usepackage{hyperref}
\usepackage{epsfig}
\usepackage{graphicx}
\usepackage{amsmath}
\usepackage{eucal}
\usepackage{latexsym}
\usepackage{amstext}
\usepackage{verbatim}
\usepackage{fancyhdr}
\usepackage{graphicx}
\usepackage{dcolumn}
\usepackage{bm}
\usepackage{hyperref}
\usepackage{amsmath}
\usepackage{amssymb}

\newcommand{\mb}[1]{ \mbox{\boldmath$#1$} }

\newcommand{\beq}{\begin{equation}}
	\newcommand{\eeq}{\end{equation}}
\newcommand{\beqq}{\begin{equation*}}
	\newcommand{\eeqq}{\end{equation*}}
\newcommand{\beqa}{\begin{eqnarray}}
	\newcommand{\eeqa}{\end{eqnarray}}
\newcommand{\p}{\partial}

\newcommand{\eps}{\varepsilon}
\newcommand{\x}{\mbox{\boldmath$x$}}

\newcommand{\y}{\mbox{\boldmath$y$}}

\newcommand{\n}{\mbox{\boldmath$n$}}

\newcommand{\pt}{\tilde{p}}
\font\bb=msbm10 at 12pt

\def\eE{\hbox{\bb E}}

\usepackage{xcolor}

\begin{document}
\title{{How large should be the redundant numbers of copy to make a rare event probable}}
\author{F. Paquin-Lefebvre$^{1}$, S. Toste$^{1}$ and D. Holcman$^{1,2}$} \affiliation{$^{1}$Group of Data modeling, Computational Biology and Applied Mathematics, Ecole Normale Sup\'erieure-PSL, 75005 Paris, France. $^{2}$  Churchill College, DAMPT, University of Cambridge, CB30DS UK.}
\begin{abstract}
The redundancy principle provides the framework to study how rare events are made possible with probability 1 in accelerated time, by making many copies of similar random searchers. But what is $n$ large? To estimate large $n$ with respect to the geometrical properties of a domain and the dynamics, we present here a criteria based on splitting probabilities between a small fraction of the exploration space associated to an activation process and other absorbing regions where trajectories can be terminated. We obtain explicit computations especially when there is a killing region located inside the domain that we compare with stochastic simulations. We present also examples of extreme trajectories with killing in dimension 2. For a large $n$, the optimal trajectories avoid penetrating inside the killing region. Finally we discuss some applications to cell biology.
\end{abstract}
\maketitle
The redundancy principle in biology expresses the need of having many redundant copies of the same particles (molecules, proteins, ions, etc...) moving randomly to trigger a rare event  such as a physiological function. The key step consists in finding a small target by the fastest particles, leading to extreme statistics \cite{weiss1983order,majumdar2016exact}. The target can be a single or a complex ensemble of molecules \cite{schuss2019redundancy,condamin2007first}. The large copy number is used to transform such a rare event, which would take a very long time compared to the other times involved in the system, into a fast event triggered by the fastest particles to arrive to the hidden target. This principle is at the basis of many cell processes such as signal transduction \cite{fain2019sensory}, cell signaling, immunology T-cell fast recognition or specific G-protein promoter selection in the nucleus \cite{alberts2013essential}. \\
In general, the arrival time of the fastest is studied by extreme statistic approaches \cite{bray2013persistence,schehr2014exact, majumdar2020extreme,lawley2020JMathB}. The redundancy theory \cite{Holcman2015,schuss2019redundancy,coombs2019first, sokolov2019extreme} applies to the molecular level but also to cells such as spermatozoa \cite{reynaud2015so}.
Yet, the theoretical computations for the time of these rare events for diffusion or anomalous diffusion \cite{lawley2020anomalous,grebenkov2020single} often relies on asymptotic approaches based on Laplace's method, when the number $n$ of players is large. But this method does not provide an order of magnitude for the number $n$. The Laplace's method and related approaches provide a formal expansion and do not allow a comparison of $n$ with physical quantities, as it assumes to begin with that $n$ is large.\\
The goal of this letter is to propose a computational framework to compare large $n$ with the scales originated from the stochastic dynamics and the small subspace configuration that defines rare events. We first present the framework to quantify large values of $n$ using the probability to find a small target. Second, to illustrate this framework, we estimate the order of magnitude for the number of copies in two cases: 1) when the search by the fastest Brownian particle has multiple choices to escape or when there is a degradation source that interferes with the exploration of the space. Interestingly, the fastest trajectories concentrate along the shortest path that can be obtained from a variational method as revealed by the Large Deviation Principle \cite{freidlin1996markov,freidlin1998random}. Interestingly for $n$ large, the optimal trajectory follows a path that avoids the killing region while staying sufficiently close.  When killing occurs in a sub-region of the domain, we obtain an explicit expression for the escape provability and an estimate of $n$. Finally, we discuss several applications in cellular biology. \\
{\bf \noindent Compensating a rare event by increasing the copy number $n$.}
To make a rare event probable when it is triggered by the arrival to a small target, one possibility is to increase the copy number $n$ of the independent identical Brownian particles. We consider here the continuous case, but the dynamics could occur on a discrete ensemble, a graph or any other topological structures, and the target is a small fraction of the parameter space. We propose the following criteria to estimate the number $n$. We consider that there are at least two fates for the Brownian particles: either they find the small target after a long time, or they get lost through another absorbing hole or are simply degraded by a killing field. When the escape probability $P_e$ is much smaller than the probability of being degraded, then the number $n$ should be large enough to compensate for the rare escape event which will occur with probability 1 for at least one, two, etc, or all particles, that is
\beq
1-(1-P_e)^n\approx 1.
\eeq
Thus when $P_e\ll1$, we propose the general elementary criteria that
\beq
n\approx \frac{1}{P_e}.
\eeq
In general the probability $P_e$ is small and almost constant far away from the narrow target window \cite{schuss2007narrow,benichou2014first,condamin2007first}. In the rest of this letter, we present specific estimates for $n$ in comparison with geometrical and dynamical characteristics.\\
\begin{figure*}[http!]
   \begin{center}
\centerline{\includegraphics[width=0.8\textwidth]{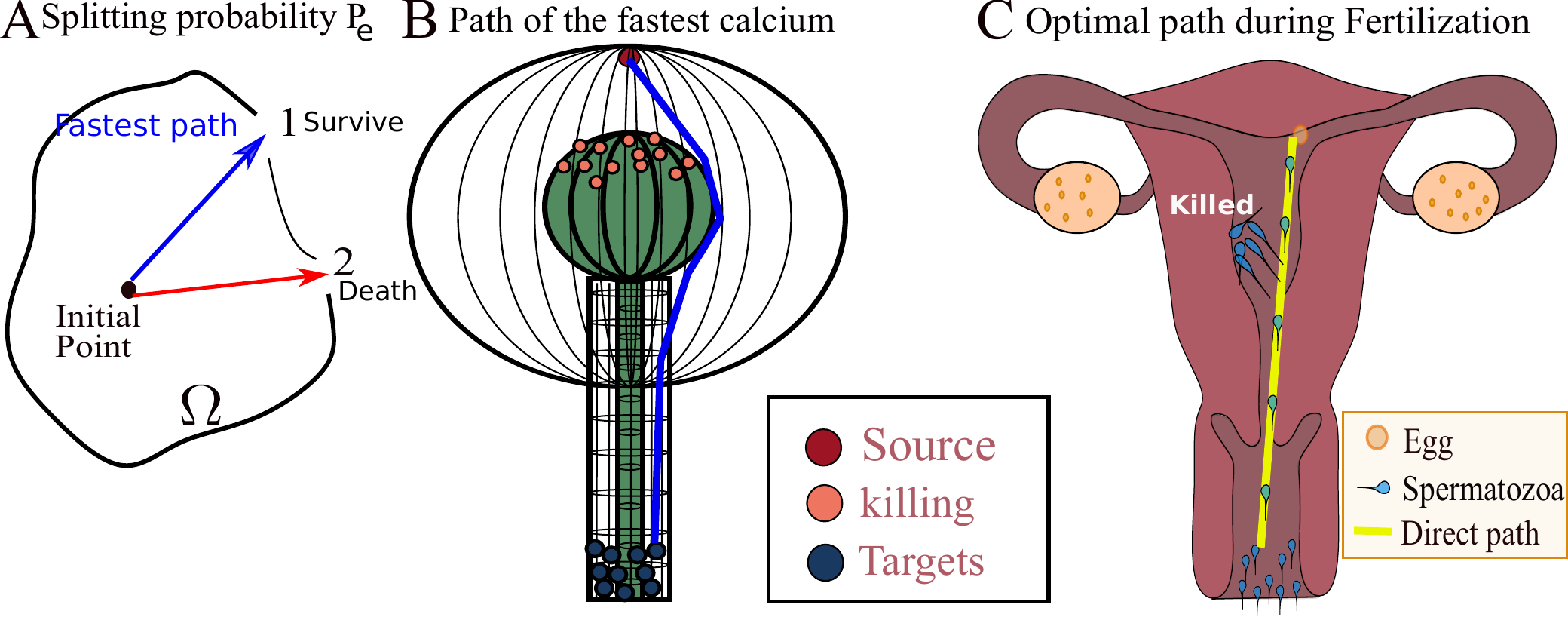}}
   \end{center}
   \caption{ {\bf Fastest escape trajectory under splitting or killing.} {\bf (A)} the fastest trajectory has the choice to escape either in 1 (relevant choice) or 2 (bad choice),.
{\bf (B)} The fastest trajectory to arrive at the base of a dendritic spine's type domain can also be absorbed on their way.
{\bf (C)} The fastest spermatozoa can arrive directly to the egg or be degraded.}
   \label{FIG1}
\end{figure*}
{\bf \noindent Computing the escape probability.}
The stochastic dynamics follows
\beq\label{eqstochastic}
\dot{\x}=\mb{b}(\x)+\sqrt{2}\mb{B}(\x)\,\dot{\mb{w}}\hspace{0.5em}\mbox{for}\ \x\in\Omega,
\eeq
where $\mb{b}(\x)$ is a smooth drift vector, $\mb{B}(\x)$ is a diffusion tensor, and $\mb{w}$ is a vector of independent standard Brownian motions. A killing field $k(\x)$ is added in the domain $\Omega$ with boundary $\p\Omega=\p\Omega_a\cup\p\Omega_r$, where $\p\Omega_a$ is a small absorbing region and $\p\Omega_r$ is reflecting.  The transition probability density function (pdf) of the process $\x(t)$ with killing is the pdf of the trajectories that have neither been killed nor absorbed in $\p\Omega_a$ by time $t$, satisfying
\beq \label{pxTtau}
 p(\x,t\,|\,\y)\,d\x=\Pr\{\x(t)\in\x+d\x,\,\tau^k>t,\,\tau^e>t\,|\,\y\},
\eeq
where $\tau^k$ (resp.~$\tau^e$) is the time for one particle to be killed (resp.~absorbed). The pdf is the solution of the Fokker-Planck equation (FPE) \cite{Schuss:Book}
\beq \label{FPEp}
 \frac{\p p(\x,t\,|\,\y)}{\p t}={\cal L}_{\x} p(\x,t\,|\,\y)-k(\x)p(\x,t\,|\,\y)\hspace{0.5em}\ \mbox{for}
\ \x,\y\in \Omega,
 \eeq
where
\beq\label{FO}
\begin{split}
&{\cal L}_{\x}p(\x,t\,|\,\y)\\
&=\sum_{i,j=1}^d\frac{\p^2\sigma^{i,j}(\x)
p(\x,t\,|\,\y)}{\p x^i\p x^j} - \sum_{i=1}^d\frac{\p b^i(\x)p(\x,t\,|\,\y)}{\p x^i}\,,
\end{split}
\eeq
and $\mb{\sigma}(\x)=\frac12\mb{B}(\x)\mb{B}^T(\x)$. The flux density vector is given by ${\cal L}_{\x}p(\x,t\,|\,\y)=-\nabla\cdot\mb{J}(\x,t\,|\,\y)$, where the components of the flux density vector are $i=1,2, \ldots, d$
\beq
J^i(\x,t\,|\,\y)=-\sum_{j=1}^d\frac{\p\sigma^{i,j}(\x)p(\x,t\,|\,\y)}{\p
x^i}+b^i(\x)p(\x,t\,|\,\y). \label{Ji1}
\eeq
The initial and boundary conditions are
\beq
p(\x,0\,|\,\y)=\,\delta(\x-\y)\mbox{ for }\
\x,\y\in \Omega\label{ICdD}
\eeq
\beq
p(\x,t\,|\,\y)=\,0 \mbox{ for } t>0 \
\x\in\p \Omega_a,\ \y\in \Omega
\eeq
\beq
\mb{J}(\x,t\,|\,\y)\cdot\n(\x)=\,0\mbox{ for }\  t>0,\ \x\in\p
\Omega-\p \Omega_a,\ \y\in \Omega. \nonumber
\eeq
{\bf \noindent Large $n$ when the splitting probability is divided into several absorbing patches}\\
In the absence of any killing processes, the large number $n$ can be estimated by the reciprocal of the splitting probability for Brownian particles that can either reach portions $\p\Omega_{a_k}$ with $k>1$ of the boundary where no action is taken, or a small boundary $\p\Omega_{a_1}$ associated with triggering a key event. The splitting probability is solution of
\beq
{\cal L*}_{\x} p(\x)=0,
\eeq
where $p(\x) =1 \hbox{ on } \p\Omega_{a_1}$ and  $p(\x) =0 \hbox{ on } \Omega_{a_k}$ with $k>1$, while $\mb{J}(\x)\cdot\n(\x)=0  \hbox{ on } \p \Omega_r$ on the reflecting part of the boundary.  The solution depends on the local geometry near the window \cite{Holcman2015}. In dimension 2, for regular absorbing windows of size $\eps_k$, the escape probability in window one is given by $ P_e= \frac{1/\log (\eps_1)}{\sum_k 1/\log (\eps_k)}$. When the windows are located at the end of a cusp with curvature $l_k$ and size $a_k$, then $P_e= \frac{\sqrt{\frac{a_1}{l_1}}}{\sum_k \sqrt{\frac{a_k}{l_k}}}$. Similar formulas in dimension 3 are $P_e= \frac{\frac{1}{\eps_1} }{\sum_k \frac{1}{\eps_k}}$ for narrow windows located on a flat boundary, and $P_e= \frac{\sqrt{\frac{a^3_1}{l_1}}}{\sum_k \sqrt{\frac{a^3_k}{l_k}}}$ when they are located at the end of a cusp. To conclude, the large number of stochastic particles depends on the size of the critical window to be reached versus the other exits. For equal size windows, the number $n$ of copies is of the order of the number of windows. \\
{\bf \noindent How large is $n$ when searching for a narrow target with a killing field}\\
We now estimate the large number $n$ when there is a killing field that can destroy particles at an exponential rate before they arrive to the target, thus leaving few particles alive.  The splitting probability is solution of the boundary value problem
 \begin{align}
\label{fdt0p} {\cal L}_{\x} \tilde p(\x\,|\,\y) -k(\x)\tilde p(\x\,|\,\y) =\, -p_I(\x)\hspace{0.5em} \hbox{for}\ \x,\y \in \Omega\\
\tilde p(\x\,|\,\y)=\,0\hspace{0.5em}\mbox{for}\ \x\in\p \Omega_a,\ \y\in \Omega\nonumber\\
\mb{J}(\x\,|\,\y)\cdot\n(\x)=\,0\hspace{0.5em}\mbox{for } \x\in\p \Omega-\p \Omega_a,\
\y\in \Omega.\nonumber
\end{align}
When the initial pdf $p_I(\x)$ is normalized ($\int_\Omega p_I(\x)d\x = 1$), the probability $P_e$ of trajectories that are terminated at $\p \Omega_a$ is given by
\beq\label{eq-pn}
P_e =\int\limits_{\Omega}\Pr\{\tau<T\,|\, {\y} \} p_I(\y)\,d\y = 1-
\int\limits_{\Omega}k(\x)\tilde{p}(\x)\,d\x.
\eeq
We now set $p_I(\x) = \delta(\x)$ and focus on the solution of
\begin{align}\label{eq:ptil}
D\Delta \tilde{p}(\x)-k(\x)\tilde{p}(\x)= -\delta(\x)\hbox{ for }\ \x \in \Omega\nonumber  \\
\tilde{p}(\x)=0\hspace{0.5em}\hbox{for}\ \x\in\p \Omega_a\nonumber  \\
\frac{\p\tilde{p}(\x)}{\p \n}=0\hspace{0.5em}\hbox{for}\ \x\in\p\Omega-\p\Omega_a,
\end{align}
where $D$ is the diffusion coefficient, and $\p\Omega_a$ is a narrow absorbing target of radius $a$ centered in $\x_a$ and $\Omega$ is a disk of radius $R$. The destruction of Brownian particles happens within a smaller disk $\Omega_\epsilon$ of radius $\epsilon$ centered in $\x_\epsilon$ (Fig.~\ref{FIG2}A) with a large constant killing rate $k(\x)$ inside this region and zero outside. The center $\x_\epsilon$ is located on the radial segment connecting the origin to the absorbing window $\p\Omega_a$.  For a large killing rate, fast particles must avoid the killing area in order to escape, as we shall see later on. The asymptotic approximation for the splitting probability is computed from $\tilde{p}(\x)$, which is itself obtained from the 2-D Neumann-Green's function $G(\x;\y)$. This function is defined as the solution of
\beq
D\Delta G(\x;\y) = \frac{1}{|\Omega|} - \delta(\x-\y)\,, \quad \x, \y \in \Omega
\eeq
with the boundary conditions: $\frac{\p G(\x,\y)}{\p\n} = 0\,, \quad \x \in \p\Omega\,, \y \in \Omega$ and $\int_\Omega G(\x;\y)d\x = 0$. In the disk of radius $R$, the solution is \cite{Cheviakov2011,ward2005,Pillay2010}
\beq
G(\x;\y) = -\frac{1}{2\pi D}\log\left(\frac{1}{R}\|\x-\y\|\right) + V(\x;\y)\,,
\eeq
where $V(\x,\y)$ is the regular part when $\|\y\| < R$, given by
\beq
\begin{split}
V(\x;\y) = &-\frac{1}{2\pi D}\left(\log\left(\left\|\frac{1}{R^2}\x\|\y\| - \frac{\y}{\|\y\|}\right\|\right) \right. \\
&\left. - \frac{1}{2R^2}\left(\|\x\|^2 + \|\y\|^2\right) + \frac{3}{4\pi}\right)\,.
\end{split}
\eeq
Using Green's identity the splitting probability can be expressed as
\begin{equation}
\begin{split}
\pt(\y) = &\frac{1}{|\Omega|}\int_{\Omega}\pt(\x) d\x + \int_{\p\Omega_a} G(\x,\y)D\frac{\p\pt}{\p\n}d\x \\
&+ G(\bm{0},\y) - k\int_{\Omega_\epsilon} G(\x,\y)\pt(\x)d\x\,,
\end{split}
\end{equation}
where $\bm{0}$ is the center of the disk. A direct computation using the absorbing boundary condition and the approximation by a constant for the solution within the killing area leads to the following expression
\begin{widetext}
\begin{eqnarray*}
\pt(\x_\epsilon) = \frac{\frac{1}{k\pi\epsilon^2}\left(\log\left(\frac{R}{a}\right) + \frac{9}{8} + \pi D\left(G(\bm{0},\x_\epsilon) - G(\bm{0},\x_a) - G(\x_a,\x_\epsilon)\right)\right)}{\frac{D}{k\epsilon^2} + \frac{1}{2}\log\left(\frac{R}{\epsilon}\right) + \log\left(\frac{R}{a}\right) + \frac{11}{8} + \pi D\left(V(\x_\epsilon;\x_\epsilon)-2G(\x_\epsilon,\x_a)\right)}.
\end{eqnarray*}
Using the relation $P_e \approx 1 - \pi \epsilon^2 k p(\x_\epsilon)$, we obtain that the splitting probability is given by
\begin{eqnarray}\label{eq:split_prob}
P_e = \frac{\frac{D}{k\epsilon^2} + \frac{1}{2}\log\left(\frac{R}{\epsilon}\right) + \frac{1}{4} + \pi D\left(V\left(\x_\epsilon;\x_\epsilon\right) - G\left(\x_a;\x_\epsilon\right) + G\left(\bm{0};\x_a\right) - G\left(\bm{0};\x_\epsilon\right)\right)}{\frac{D}{k\epsilon^2} + \frac{1}{2}\log\left(\frac{R}{\epsilon}\right) + \log\left(\frac{R}{a}\right) + \frac{11}{8} + \pi D\left(V\left(\x_\epsilon;\x_\epsilon\right) - 2G\left(\x_a;\x_\epsilon\right)\right)}\,.
\end{eqnarray}
\end{widetext}
This formula reveals the role of each parameter and in particular, for $\frac{1}{k\epsilon^2}$ large, we get
\beq
P_e \approx \frac{\frac{D}{k\epsilon^2} + \frac{1}{2}\log\left(\frac{R}{\epsilon}\right)}{\frac{D}{k\epsilon^2} + \frac{1}{2}\log\left(\frac{R}{\epsilon}\right) + \log\left(\frac{R}{a}\right)}.
\eeq
To study the range of validity of the asymptotic solution \eqref{eq:split_prob}, we use stochastic simulations for 1000 runs with particles starting at the origin of a disk of radius $R=2$. The discrete time step is $\Delta t = 0.001$ and a parameter sweep of the killing rate is performed over the range $k = [1\,\,\,10\,\,\,25\,\,\,50\,\,\,75\,\,\,90]$. The other geometrical parameters are as indicated in the caption of Fig.~\ref{FIG2}. We also compute the splitting probability $P_e$ directly by solving the system \eqref{eq:ptil} using the finite element solver COMSOL \cite{comsol}, for which we approximate the killing field by a smooth function $k(\x) \equiv \frac{k}{2}\left(1 + \tanh\left(\frac{\epsilon - \|\x-\x_\epsilon\|}{\Gamma}\right)\right)\,$, where $\Gamma=0.0001 \ll \epsilon$. Here, $\Gamma$ is a parameter that controls the width of the transition layer. The 2-D Dirac delta function $\delta(\x)$ is approximated by a 2-D Gaussian $p_I(\x) = \frac{1}{2\pi\sigma^2}\exp{\left(-\frac{\x^T\x}{2\sigma^2}\right)}\,$ with a small variance $\sigma=0.01 \ll 1$. We obtain a good agreement between the numerical and asymptotic solutions both when the killing region is located inside or tangent to the boundary (Fig.~\ref{FIG2}B).
\begin{figure*}[http!]
\begin{center}
\centerline{\includegraphics[width=1\textwidth]{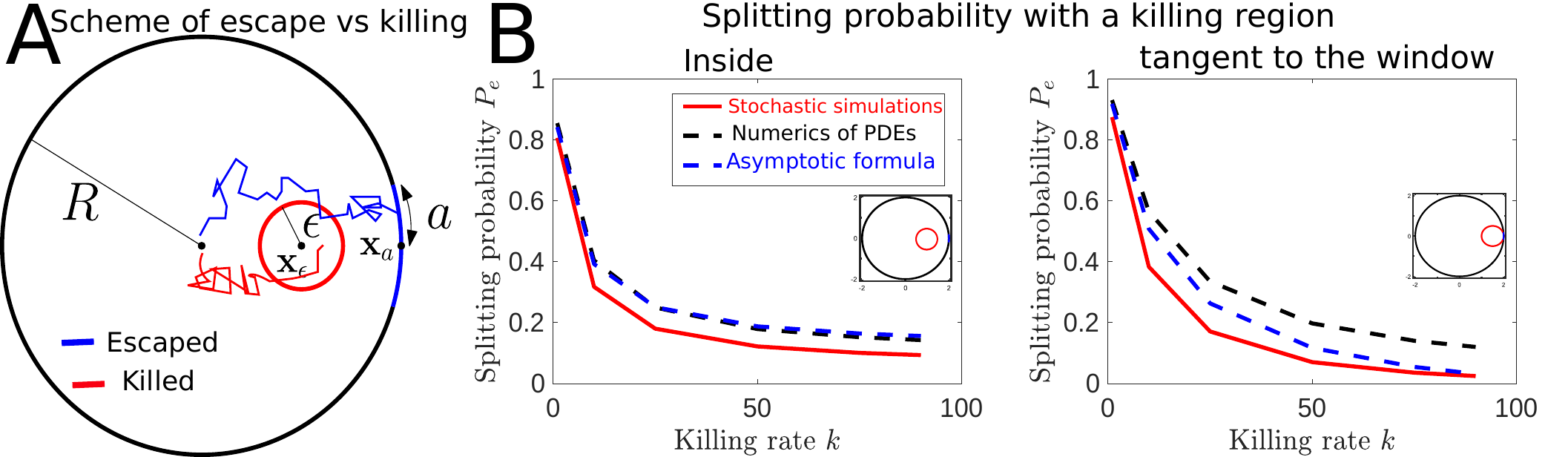}}
\caption{{\bf Splitting probability versus killing.} {\small {\bf (A)} Scheme of the domain: a disk $\Omega$ of radius $R$ containing a killing region $\Omega_\epsilon$ (red) of size $\epsilon$. Particles starting at the center can either escape through a window $\p\Omega_a$ (blue) of narrow radius $a$ and centered in $\x_a$ and can be terminated (red) inside the domain $\Omega_\epsilon$. {\bf (B)} Splitting probability versus the killing rate $k$ when the killing region is located inside (Left) and tangent  (Right) to the absorbing window for stochastic simulations (red), the asymptotic solution \eqref{eq:split_prob} (dashed-blue) and the numerical solution of the PDE equation \eqref{eq:ptil}(dashed-black). The parameters are $R=2$, $D=1$, $\epsilon = 0.25$, $a = 5\pi R/180$, $\x_a=(R,0)^T$, and $\x_\epsilon = (1,0)^T$ (Left) or $\x_\epsilon = (1.75,0)^T$ (Right).}}
\label{FIG2}
\end{center}
\end{figure*}
To conclude, the number of redundant copies should be of the order of
\beq
n \approx \frac{\frac{D}{k\epsilon^2} + \frac{1}{2}\log\left(\frac{R}{\epsilon}\right) + \log\left(\frac{R}{a}\right)}{\frac{D}{k\epsilon^2} + \frac{1}{2}\log\left(\frac{R}{\epsilon}\right)}\,,
\eeq
which decreases with the radius $a$ of the absorbing window, but also depends on the killing parameters. Note that this formula is valid for $k$ fixed with $\epsilon$ tending to zero.\\
{\bf \noindent Optimal exit paths for the fastest with a killing field}\\
In the absence of a killing field, according to the Large Deviation Principle \cite{freidlin1996markov}, the fastest particles use a path toward the absorbing window well concentrated near the shortest geodesic as the number of particles increase. With a killing region, we expect a deviation such that the shortest geodesic should avoid this area when the killing rate is large. To estimate this optimal path of the fastest, we use arguments from the Large Deviation Principle. We recall that the fluctuations of the diffusion process with small amplitude noise $D$ around the deterministic function $F$ is given by the action functional in the time interval $[0,T]$, \cite{martin2019long,schehr2014exact,majumdar2020extreme}
\beq \label{large}
\text{Pr } \left\{\max_{t\in [0,T]}|x(t)-F(t)| \right\} \asymp \exp \left\{ -\frac{S_{F}(T)}{D}\right\},
\eeq
where $x$ is a Brownian motion with diffusion $D$ and zero mean and
\beq
S_{F}(T)=\frac{1}{4} \int_{0}^{T} |\dot F(s)|^2ds.
\eeq
To account for the killing field $k(x)$, we use the Feynman-Kac representation for the solution of the FPE
\beq \label{FKformula}
p(x,T) =\eE\left[ f(x(s))\exp \left(-\int_{0}^{T} k(x(s))ds\right) \right],
\eeq
where $x$ starts at time $0$ at position $\x$ and $f$ is the initial distribution equal to the delta Dirac at 0. The survival probability is given by
\begin{widetext}
\beq
\int_{0}^{\infty} \left(\int_{\Omega} p(x,t)dx\right)^n dt \approx  \int_{0}^{\infty}  \int_ {\{ \hbox{ x(s) paths from }0 \hbox{ to } \p\Omega_a \}}\exp \left\lbrace\frac{-n}{4D} \int_{0}^{t} |\dot x(s)|^2ds -n\left(\int_{0}^{t} k(x(s))ds\right)\right\rbrace d\mathcal{D}(x(s)) dt.
\eeq
\end{widetext}
The contribution of the integral for large $n$ occurs at the minimum of the functional:
\beq
\underset{\underset{X(T)\in\p \Omega_a\}}{\{x(0)=x,} }{\min} \left\lbrace \frac{1}{4D} \int_{0}^{t} |\dot x(s)|^2ds+\left(\int_{0}^{t} k(x(s))ds\right)\right\rbrace.
\eeq
Applying Euler's Lagrange principle, we obtain that the minimum occurs along the optimal trajectory, solution of the differential equation:
\beq\label{Euler}
-\ddot{x}+2 D\nabla k(x(s))=0,
\eeq
with $x(0)=x$ and $x(T) \in \p \Omega_a$. As shown by the Brownian simulations, the trajectories of the fastest particles starting from the origin are concentrated along a path solution of equation \eqref{Euler}. The optimal paths avoid the killing region as the killing rate increases (Fig.~\ref{FIG3}A). Interestingly, as the killing region become tangent to the absorbing window, the density of the fastest spreads around the origin and also inside the entire domain (yellow) (Fig.~\ref{FIG3}B). To conclude, the fastest trajectories avoid the killing region.\\
\begin{figure*}[http!]
\begin{center}
\centerline{\includegraphics[width=0.9\textwidth]{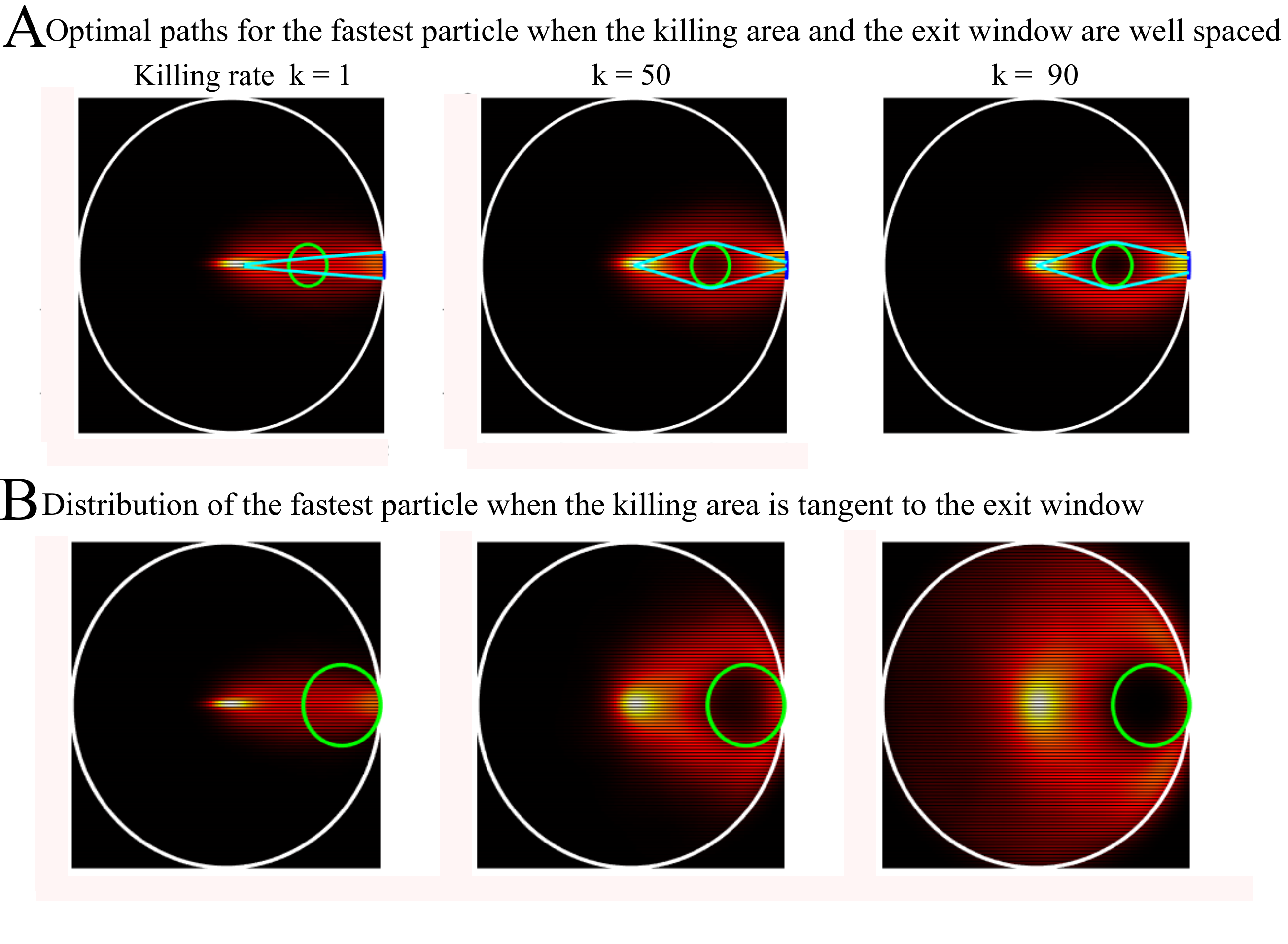}}
\caption{ {\bf Distribution of the trajectories for the fastest associated to two positions of the killing region} {\small {\bf (A)}
inside and {\bf (B)} tangent to the absorbing window. The trajectory solutions of \eqref{Euler} (cyan) are computed for the case (A), but not (B) since the variational problem does not account for possible bouncing rays. The parameters are $R=2$, $D=1$, $a = 5\pi R/180$, $\x_a=(R,0)^T$, and $\x_\epsilon = (1,0)^T$, $\epsilon = 0.25$ {\bf (A)} or $\x_\epsilon = (1.5,0)^T$, $\epsilon = 0.5$ {\bf (B)}. }}
\label{FIG3}
\end{center}
\end{figure*}
{\bf \noindent Discussion and concluding remarks}\\
Multiplying the copy number of random molecules, ions, proteins or cells is a key process to make a rare event frequent. It is at the basis of the redundancy principle \cite{schuss2019redundancy}. Although the principle acknowledges that a large number is needed, it does not specify the order of magnitude. In this letter, we presented a computational framework to estimate the number of copies to guarantee that the rare event will at least occur with one arrival particle. The rare event is described by the splitting probability to find the activation before escaping through a much more probable place, or before being degraded. Explicit and asymptotic computations of the probability allow to determine the role of the geometry and the dynamics in the estimation of the large number $n$ of copies.\\
In the process of neuronal transmission at synapses, the splitting probability between small receptors located on the post-synaptic region and the lateral opening of the synaptic cleft is of the order of $P_{act} \approx 10^{-3}$. Interestingly, the number of neurotransmitters is of the order of few thousands (2000 to 3000) \cite{kandel2000}. Modulating this number is key for controlling neurotransmission, and a reduction by 20\% \cite{nicoll2010dietEdwards} resulting from a ketogenic diet can stop some epileptic crisis. But the most classical example is certainly the fertility process, where hundreds of millions of spermatozoa are required to guarantee that fertilization can be possible: the large number compensates for the long distance, which cannot be recovered by other mechanisms such as chemotaxis, rheotaxis or thermotaxis \cite{kaupp2017signaling}. Here the large redundancy is particularly required, as a decline by 10-20\% is associated to infertility. Finally, it would be interesting to generalize the present large $n$ estimation approach to different random motions (fractional Brownian motion, Levy flight), or any other anomalous diffusion processes.
\normalem
\bibliographystyle{apsrev4-1}

\bibliography{PRL_large_n}
\end{document}